\documentstyle[multicol,aps,prl]{revtex}

\draft

\begin{document}
\draft
\title{N-qubit Entanglement Index and Classification}
\author{Ying Wu}
\address{Physics Department and National Key Laboratory for
Laser Technique, Huazhong University of Science and Technology,
Wuhan 430074, and Wuhan Institute of Physics and Mathematics, The
Chinese Academy of Sciences, Wuhan 430071, People's Republic of
China}
\date{\today}
\maketitle
\begin{abstract}
We show that all the N-qubit states can be classified as N
entanglement classes each of which has an entanglement index
$E=N-p=0,1,\cdots,N-1$($E=0$ corresponds to a fully separate
class) where $p$ denotes number of groups for a partition of the
positive integer N. In other words, for any partition
$(n_1,n_2,\cdots,n_p)$ of N
 with $n_j\ge 1$ and $N=\sum_{j=1}^{p}n_j$,
the entanglement index for the corresponding state
$\rho_{n_1}\bigotimes\rho_{n_2}\cdots\bigotimes \rho_{n_p}$ with
$\rho_{n_j}$ denoting a fully entangled state of $n_j-$qubits is
$E(\rho_{n_1}\bigotimes\rho_{n_2}\cdots\bigotimes
\rho_{n_p})=\sum_{j=1}^{p}(n_j-1)=N-p$.
\end{abstract}
\begin{multicols}{2}
\maketitle

Recently, Yu, {\it et al.} have classified N-qubit entanglement
via Bell's inequalities as $(N-1)$ entanglement classes with an
entanglement index $E=0,2,3,\cdots,N$\cite{s1}. Although their
original idea is elegant, their classification and introduced
entanglement index suffer from serious drawbacks. For instance, it
is easily checked to find that their classification and
entanglement index can not, even for a 2-qubit system,
discriminate the fully separable 2-qubit states such as
($|0\rangle_{1}|0\rangle_{2}$) from the any Bell states (such as
($|0\rangle_{1}|0\rangle_{2}+|1\rangle_{1}|1\rangle_{2}$) which
are maximum entangled states for a 2-qubit system, and some of
their  classification and entanglement index do not satisfy the
additivity feature for tensor products of independent states,
which should be satisfied by any correct entanglement
measure\cite{s4}. These mentioned unreasonable features may
originate from the fact\cite{s2,s3} that the CHSH
(Clauser-Horne-Shimony-Holt) form of Bell's inequalities are not a
sufficiently good measure of quantum correlations in the sense
that there are states which don't violate the CHSH inequality but
on the other hand can be purified by local interactions and
classical communications to yield a state that does violate the
CHSH inequality.

Closely inspecting the unreasonable features mentioned above, we
find that they can be removed by introducing a new scheme of
classification and entanglement index as follows: All the N-qubit
states can be classified as N entanglement classes each of which
has an entanglement index $E=N-p=0,1,\cdots,N-1$ ($E=0$
corresponds to a fully separate class). Here $p$ denotes number of
groups for a partition of the positive integer N, i.e., the
partition of N ( $(n_1,n_2,\cdots,n_p)$ with $n_j\ge 1$ and
$N=\sum_{j=1}^{p}n_j$) corresponds a set of p strictly positive
integers that sum up to N.

Let us explain this novel scheme in some details. All the
partitions for a fixed positive integer N can be easily obtained
by a simple Mathematica code "$>>$ DiscreteMath`Combinatorica` ;
n=specified positive integer; Partitions[n]". Obviously, there
exists a one-to-one correspondence\cite{s1}  between a partition
$(n_1,n_2,\cdots,n_p)$ with $n_j\ge 1$ and $N=\sum_{j=1}^{p}n_j$
and a N-qubit state $
\rho_{n_1}\bigotimes\rho_{n_2}\cdots\bigotimes \rho_{n_p}$ with
$\rho_{n_j}$ denoting a fully entangled state of $n_j-$qubits or
called a $n_j$-qubit GHZ state \cite{s4} (Actually, $n_j\ge 3$
corresponds to a real GHZ state while $n_j=2$ denotes a fully
entangled state or a EPR state while $n_j=1$ does not correspond
to any entangled state). The entanglement index for such a N-qubit
state is defined as $E(n_1,n_2,\cdots,n_p)\equiv
E(\rho_{n_1}\bigotimes\rho_{n_2}\cdots\bigotimes
\rho_{n_p})=\sum_{j=1}^{p}E(\rho_{n_j})$. The entanglement index
for any $n_j$-qubit GHZ state (including the 1-particle states
corresponding to $n_j=1$) can be defined as $ E(\rho_{n_j})\equiv
E(n_j)=(n_j-1)$ and hence
$E(n_1,n_2,\cdots,n_p)=\sum_{j}^{p}E(n_j)=N-p$. It is pointed out
that such a specification of entanglement index for the  any
$n_j$-qubit GHZ state is consistent with the conclusion\cite{s4}
that a m-qubit GHZ state can be made from a set of (m-1) EPR pairs
corresponding to $E(m)=(m-1) E(2)=m-1$ in our notation. Then by
noting that the positive integer p for a given positive integer N
can take all the integers $1,2,\cdots,N$, we see that the
entanglement index for a fixed N is $E(n_1,n_2,\cdots,n_p)\equiv
E(\rho_{n_1}\bigotimes\rho_{n_2}\cdots\bigotimes
\rho_{n_p})=\sum_{j=1}^{p}E(\rho_{n_j})=\sum_{j=1}^{p}(n_j-1)=N-p$
which can take the values $E=N-p=0,1,\cdots,N-1$.
$E=1,2,\cdots,(N-1)$ corresponds to $(N-1)$ different entangled
classes of a N-qubit system while $E=0$ corresponds to a fully
separate class.

Such a scheme of classification and entanglement index obviously
represents a reasonable one because it can readily seen that it
satisfies the following four properties for any reasonable
entanglement measure\cite{s4}: 1)It should be zero for separate
states (our scheme satisfies this requirement because
$E(1,1,\cdots,1)=\sum_{j=1}^{N}E(1)\equiv 0 $ due to
$E(1)=1-1=0$). 2)It should be invariant under local unitary
transformations (our scheme satisfies this property because
$E(U_{local}\rho_{n_1}\bigotimes\rho_{n_2}\cdots\bigotimes
\rho_{n_p}U_{local}^{\dag})=\sum_{j=1}^{p}E(U_{j}\rho_{n_j}
U_{j}^{\dag})=\sum_{j=1}^{p}E(\rho_{n_j})$ where
$U_{local}=\prod_{j}^{p}U_{j}$ with $U_{j}$ denoting any unitary
operation on the specified $n_j-$qubit system). 3) Its expectation
should not increase under local quantum operations and classical
communication (this property is obviously satisfied by our
scheme). 4)It should be additive for tensor products of
independent states, shared among the same set of observers (our
scheme satisfies obviously for any partition with $p>1$ while for
$p=1$ it is also true due to the conclusion\cite{s4} that a
m-qubit GHZ state can be made from a set of (m-1) EPR pairs
corresponding to $E(m)=(m-1) E(2)=m-1$ in our notation).

In summary, inspired by the elegant idea in ref.\cite{s1}, we have
proposed a simple and reasonable novel scheme classifying and
measuring the entanglement for a N-qubit system by showing that
all the N-qubit states can be classified as N entanglement classes
each of which has an entanglement index $E=N-p$ which takes one of
the values $0,1,\cdots,N-1$ ($E=0$ corresponds to a fully separate
class). Here $p$ denotes number of groups for a partition of the
positive integer N. A larger value of the entanglement index
corresponds to a bigger entanglement in the fact that the
entanglement is a kind of quantum resource with the entanglement
of any EPR pairs as a basic unit. Any n-qubit GHZ state can be
transformed into (n-1) EPR pairs by LOCC according to Bennett et
al's conclusion\cite{s4}.  It is emphasized that the entanglement
index and classification are suitable for pure states and for
mixed states as well because the entanglement index with a fixed N
for the state density operator
$\rho=\sum_{all\;p}P(n_1,n_2,\cdots,n_p)\rho_{n_1}\bigotimes\rho_{n_2}\cdots\bigotimes
\rho_{n_p}$ with the probabilities $P(n_1,n_2,\cdots,n_p)$
satisfying $\sum_{all\;p}P(n_1,n_2,\cdots,n_p)=1$  can be defined
as $E(\rho)=\sum_{all\;p}P(n_1,n_2,\cdots,n_p)
E(\rho_{n_1}\bigotimes\rho_{n_2}\cdots\bigotimes
\rho_{n_p})=\sum_{all\;p}P(n_1,n_2,\cdots,n_p)
\sum_{j=1}^{p}E(\rho_{n_j})$ or $E(\rho)=
=\sum_{all\;p}P(n_1,n_2,\cdots,n_p)(N-p)$.

YW is supported by the National Natural Science
Foundation of China through grants 90108026, 60078023 and
10125419, and by the Chinese Academy of Sciences through the 100
Talents Project and grant KJCX2-W7-4.

\end{multicols}

\end{document}